\documentclass[prl,twocolumn,showpacs,preprintnumbers,amsmath,amssymb,nofootinbib]{revtex4}
\usepackage{subfigure}
\usepackage{graphicx}
\usepackage{bm}

\newcommand{\mplanck}{M_{\rm P}}
\newcommand{\rd}{\textrm{d}}
\newcommand{\ee}{\begin{equation}}
 \newcommand{\eee}{\end{equation}}
\newcommand{\ea}{\begin{eqnarray}}
 \newcommand{\eea}{\end{eqnarray}}

\newcommand{\om}{\Omega_{\rm m}}
\newcommand{\od}{\Omega_{\rm d}}
\newcommand{\ob}{\Omega_{\rm b}}
\newcommand{\og}{\Omega_\gamma}
\newcommand{\ore}{\Omega_{\rm rel.}}
\newcommand{\olam}{\Omega_{\Lambda}}

\newcommand{\ode}{\Omega_{\rm d}^e}
\newcommand{\omh}{\Omega_{\rm m}^0 h^2}
\newcommand{\obh}{\Omega_{\rm b}^0 h^2}

\newcommand{\crest}{c_{s}^2}

\preprint{HD-THEP-06-01}
\begin{document}

\author{Michael Doran}
\email{M.Doran@thphys.uni-heidelberg.de}
\affiliation{Institut f\"ur  Theoretische Physik, Philosophenweg 16, 69120 Heidelberg, Germany}
\author{Georg Robbers}
\email{G.Robbers@thphys.uni-heidelberg.de}
\affiliation{Institut f\"ur  Theoretische Physik, Philosophenweg 16, 69120 Heidelberg, Germany}
\title{Early Dark Energy Cosmologies}
\begin{abstract}
We propose a novel parameterization of the dark energy density. It is particularly well suited
to describe a non-negligible contribution of dark energy at early times and contains
only three parameters, which are all physically meaningful: the fractional dark energy density today, the
equation of state today and the fractional dark energy density at early times.
As we parameterize $\od(a)$ directly instead of the equation of state, we can give analytic
expressions for the Hubble parameter, the conformal horizon today and at last scattering,
the sound horizon at last scattering, the acoustic scale as well as the luminosity distance.
For an equation of state today $w_0 < -1$, our model crosses the cosmological constant
boundary. We perform numerical studies to constrain the parameters of our model by
using Cosmic Microwave Background, Large Scale Structure and Supernovae Ia data.
At $95\%$ confidence, we find that the fractional dark energy density at 
early times $\ode < 0.06$. This bound tightens considerably to $\ode < 0.04$ when the
latest Boomerang data is included.
We find
that both the gold sample of Riess et. al. and the SNLS data of Astier et. al. 
when combined with CMB and LSS data mildly prefer
$w_0 < -1$, but are well compatible with a cosmological constant.
\end{abstract}

\maketitle

%%%%%%%%%%%%%%%%%%%%%%%%%%%
%%%%%%%%%%%%%%%%%%%%%%%%%%%%
%%%%%%%%%%%%%%%%%%%%%%%%%%%%
\section{Introduction}
Current observations 
\cite{Astier:2005qq,Riess:2004nr,Spergel:2003cb,Readhead:2004gy,Dickinson:2004yr,Tegmark:2003ud} favor some form of dark energy that today comprises roughly 70\% of the energy density
of our Universe. One fundamental issue is whether dark
energy is a true cosmological constant or time evolving \cite{Wetterich:fm,Ratra:1987rm,Caldwell:1997ii}.
In recent years, the notion of an evolving dark energy
has been cast in various parameterizations 
\cite{Cooray:1999da,Huterer:2000mj,Corasaniti:2002vg,Linder:2002et,Chevallier:2000qy,Upadhye:2004hh,Wang:2004py} of the 
equation of state $w(a)=p/\rho$ of dark energy. Yet such parameterizations
are ill-suited to catch an intriguing possible feature of
dark energy, namely that it could be present at an observable
level even from as early as Big Bang Nucleosynthesis on.  Such models 
would leave their imprints on the Cosmic Microwave Background \cite{Doran:2000jt},
cosmic structure \cite{Doran:2001rw,Bartelmann:2005fc} and maybe even
Big Bang Nucleosynthesis \cite{Wetterich:1994bg,Bean:2001wt,Muller:2004gu}.
There are some good
points in favor of this scenario:  if one links the presently small energy
density of dark energy to the age of the Universe, one is
 led to attractor solutions \cite{Wetterich:fm,Ratra:1987rm}.  
Such scenarios occur in attempts to solve the 
cosmological constant problem from the point of 
view of dilatation symmetry \cite{Wetterich:fm} and also in string theories.

Instead of parameterizing $w(a)$, we parameterize $\od(a)$ directly.
This will prove advantageous for two reasons: firstly, the amount of
dark energy at early times is then a natural parameter and not inferred by
integrating $w(a)$ over the entire evolution. Secondly, since the Hubble parameter
is given by
\ee\label{eqn::hubble}
\frac{H^2(a)}{H^2_0} = \frac{\om^0 a^{-3} + \ore^0 a^{-4}}{1 - \od(a)},
\eee
a simple, analytic expression for $\od(a)$ enables us to compute many astrophysical quantities
analytically. In the above,  $\ore^0$ is the fractional energy density of relativistic neutrinos and photons today,
$\om^0$ is the matter (dark and baryonic) fractional energy density and we assumed a flat Universe.
For any parameterization of $\od(a)$, the equation of state can of course be
inferred from $\od(a)$ via an analytic relation (see Equation \eqref{eqn::connectw}). 
\begin{figure}
\includegraphics[scale=0.30] {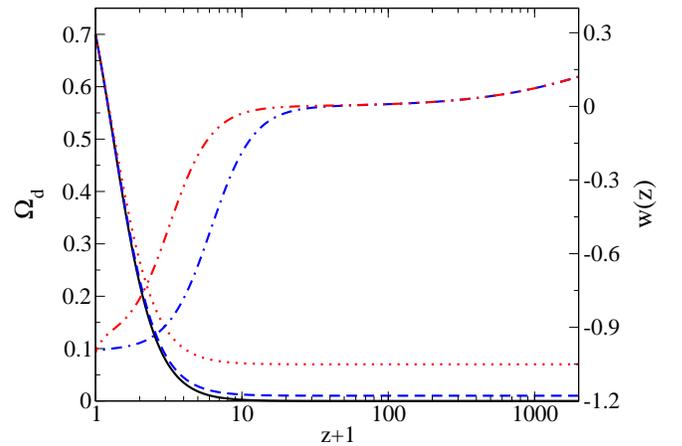}
\caption{Evolution of the fractional dark energy density $\od(z)$ and the
equation of state $w(z)$ as a function of
redshift. The solid (black) curve depicts the behavior for a $\Lambda$CDM cosmological
constant model, in which $w_0 =-1$ by definition and the amount of dark energy
at early times tends to zero. In contrast, the dashed (blue) and dotted (red)
curves are early dark energy models described by our parameterization
\protect\eqref{eqn::param}. For the models depicted, we chose $w_0=-1$ and 
$\ode= 0.01$ (blue, dashed), $\ode=0.07$  (red, dotted) respectively.
In addition, we plot the equation of state $w$ for $\ode=0.01$ (blue, dashed-dotted) 
and $\ode=0.07$ (red, dashed-double-dotted) 
\label{fig::evolution}
}
\end{figure}
\section{Parameterizing Dark Energy}
As said, we would like to parameterize $\od(a)$ to catch the important
feature of early dark energy. In addition, the parameterization
must involve only a restricted number of - physically meaningful -
parameters (we will only require three: $\od^0$, $w_0$ and $\ode$, see below).
Equation \eqref{eqn::param} is our proposal which
is derived from the following simple considerations: we start by
observing that neglecting radiation, the fractional energy density of a cosmological
constant evolves as
\ee
\olam(a) = \frac{\od^0}{\od^0 + \om^0 a^{-3}}.
\eee
Here, $a$ is the scale factor normalized to $a=1$ today, $\od^0$ and $\om^0$ are
the fractional densities of dark energy and dark matter today and we assume a flat universe, i.e.
 $\om^0 + \od^0 = 1$.
The first generalization of this formula is to allow $w \neq -1$, which 
is achieved by \cite{Doran:2001rw}
\ee
\od(a) = \frac{\od^0}{\od^0 + \om^0 a^{3w_0}}.
\eee
A straightforward attempt to include early dark energy is 
to simply add a term that is switched on at high redshifts and
gives a basic contribution of the requested level 
\ee
\od(a) =  \frac{\od^0}{\od^0 + \om^0 a^{3w_0}} + \ode (1- a^\alpha).
\eee
where $\alpha > 0$ is a parameter. It turns out, however,
that this is insufficient, because
the evolution of $\od$ is connected to the equation of state $w$
by the relation \cite{Wetterich:2004pv}
\ee \label{eqn::connectw}
\left\{3 w - \frac{a_{eq}}{a + a_{eq.}} \right\} \od (1-\od) = - \rd \od / \rd \ln a,
\eee
where $a_{eq}$ is the scale factor at matter-radiation equality.
Demanding that $w(a=1) = w_0$, i.e. that the parameter $w_0$
should indeed have its usual meaning, one is led to conclude
that an additional term is needed  in the numerator and that $\alpha=-3 w_0$.
This yields  the final form of our  parameterization, namely
\ee\label{eqn::param}
\od(a) =  \frac{\od^0 - \ode \left(1- a^{-3 w_0}\right) }{\od^0 + \om^0 a^{3w_0}} + \ode \left (1- a^{-3 w_0}\right).
\eee

In terms of the equation of state, going from today to the past, $w$ starts at $w_0$. 
It then crosses over to $w\approx 0$ during the matter dominated era. Defining
the cross-over as $w = w_0/2$ and 
using Equations  \eqref{eqn::connectw}  and \eqref{eqn::param} and working 
to leading order in $\ode/\od^0$, one obtains 
the cross-over scale factor
\ee\label{eqn::crossover}
a_c \approx \left(\frac{\od^0}{[1-\od^0]\ode}\right)^{\frac{1}{3w_0}}.
\eee 
We see from Equation \eqref{eqn::crossover}, that increasing $\ode$ increases
the cross-over scale factor, as expected. Likewise, a more negative $w_0$ also 
increases $a_c$, i.e. cross-over of $w$ occurs more recently.
Finally, in the radiation dominated era, $w$ tends to $1/3$.

It is worth mentioning that our parameterization \eqref{eqn::param} is a monotonic function
of $a$ for as long as 
\ee
\ode \lesssim \frac{\od^0}{2-\od^0},
\eee 
i.e. it stays monotonic even for rather large $\ode$. Hence, as $w=0$ during
matter domination, we see that for \mbox{$w_0 < -1$}, $w(a)$ will cross the
cosmological constant boundary \cite{Feng:2004ad,Vikman:2004dc,Hu:2004kh,Huey:2004jz,Caldwell:2005ai}. 
The evolution of $\od$ and the corresponding $w$ are depicted in Figure \ref{fig::evolution}.

\begin{figure}
\includegraphics[angle=-90,width=0.48\textwidth]{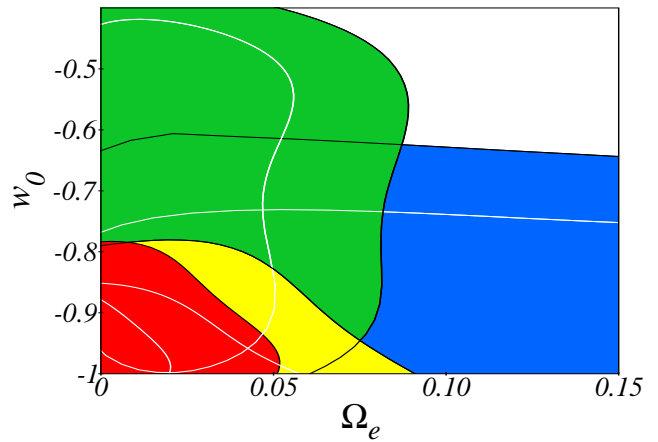}
\caption{Constraints on the parameters $w_0$ and $\ode$ for different combinations of data sets.
         The blue region corresponds to the SNe Ia compilation of Riess et. al., the green region
         to WMAP + VSA + CBI + SDSS. The
         constraints obtained when combining all of these sets are shown in yellow. The result of adding
         the Boomerang data to this combined set is depicted in red. Black (white) lines enclose 95\%  (68\%) confidence regions.
         }
\label{fig::combined_scalar_w0_Oe}
\end{figure}

\section{Analytic results}
A direct parameterization of $\od(a)$ removes one integration otherwise
necessary to compute the luminosity distance, sound horizon etc., because
the Hubble parameter $H$ is given by Equation \eqref{eqn::hubble}. Many
cosmological quantities can then be computed analytically; the 
luminosity distance $d_L$, the horizon today
and at last scattering $\tau_0$ and $\tau_{ls}$, the sound horizon $r_s$ and the acoustic
scale $l_A$. 

To derive an expression for the horizon today $\tau_0$, 
we neglect radiation in Equation \eqref{eqn::hubble}, which
leads to an error of less than $1\%$. Combining Friedmann's Equation today and at
arbitrary scale factor yields
\ee
\left(\frac{\rd a}{\rd \tau}\right)^2 = H_0^2 \left (\om^0 a + \od(a) \right).
\eee
Inverting, drawing the root and separating the variables, one gets
\ee
\rd \tau = H_0^{-1} \int \rd a\, \sqrt{\frac{1 - \od(a)}{\om^0\,a}}.
\eee
Using our parameterization Equation \eqref{eqn::param} and substituting $y=a^{-3 w_0}$,
one obtains
\ee\label{eqn::tauint}
\tau_0 = \frac{-1}{3w_0H_0\sqrt{\om^0}} \int_0^1 \frac{ y ^{-1 - \frac{1}{6w_0} }   \sqrt{1 - \ode \left[ 1 - y \right ]^2  } } {    \sqrt{1 + \frac{\od^0}{\om^0} y }} \rd y.
\eee
Expanding the root in the numerator yields
\ee
\tau_0 = \frac{-1}{3w_0H_0\sqrt{\om^0}} \int_0^1 \frac{ y ^{-1 - \frac{1}{6w_0} }   \left (1 - \frac{\ode}{2} \left[ 1 - y \right ]^2  \right) } {    \sqrt{1 + \frac{\od^0}{\om^0} y }} \rd y.
\eee
We then split the integral in leading and next to leading order
\begin{multline}
\tau_0 =  \frac{-1}{3w_0H_0\sqrt{\om^0}} \Biggl \{ \int_0^1 \frac{y ^{-1 - \frac{1}{6w_0} }    } {    \sqrt{1 + \frac{\od^0}{\om^0} y }}\rd y \\ - \frac{\ode}{2}  \int_0^1 \frac{  y^{-1 -\frac{1}{6w_0}}[1-y]^2   } {    \sqrt{1 + \frac{\od^0}{\om^0} y }}\rd y \Biggr \}.
\end{multline}
Both integrals yield hypergeometric functions $_2F_1$  (up to  $\Gamma$ functions).
The final result may hence be expressed as
\begin{multline}
\tau_0 =  \frac{2}{H_0\sqrt{\om^0}} \Biggl\{\,  _2F_1\left(\frac{1}{2},\frac{-1}{6w_0},1-\frac{1}{6w_0},-\frac{\od^0}{\om^0}\right)  \\
 - \frac{\ode}{(2-\frac{1}{6w})(1-\frac{1}{6w})}\,  _2F_1\left(\frac{1}{2},-\frac{1}{6w_0},3-\frac{1}{6w_0},-\frac{\od^0}{\om^0}         \right)  \Biggr\}.
\end{multline}
In the limit that $\ode \to 0$, we recover the result for constant $w$ \cite{Doran:2001rw}.
The luminosity distance is computed in quite the same manner. Neglecting radiation, it is given by
\begin{eqnarray}\label{eqn::dl}
d_L(z) &=& (1+z) \int_0^z \frac{\rd z'}{H(z)} = (1+z) \int_{a(z)}^{1} \frac{a^{-2} \rd a}{H(a)} \nonumber \\
&=&  \int_{a(z)}^{1} \frac{a^{-1/2}\rd a}{H_0 \sqrt{\om^0}} \sqrt{ \frac{ 1 - \ode[1-a^{-3w_0}]^2 }{ 1 + \frac{\od^0}{\om^0} a^{-3w_0}  }  }.
\end{eqnarray}
The integral is in fact identical to the one in \eqref{eqn::tauint}, but with $a(z)$ instead of $0$ as 
lower boundary.  It seems feasible to derive an expression for $d_L(z)$ in terms of hypergeometric
functions as well \cite{Gruppuso:2005xy}. Yet such an expression would be rather lengthy. As the evaluation necessitates numerical methods in any case, we leave Equation \eqref{eqn::dl} as it is.

At high redshift, $\ode$ simply scales the Hubble parameter by a constant factor.
The Friedmann Equation (this time including radiation) then yields 
the horizon at last  scattering \cite{Doran:2000jt}
\ee
\tau_{ls} = \frac{2}{H_0} \left(\frac{1-\ode}{\om^0}\right)^{1/2} \left [  \sqrt{a_{ls} + \frac{\ore}{\om^0}} - \sqrt{\frac{\ore}{\om^0}} \right].
\eee
Here, $a_{ls}$ is the scale factor at recombination.

The sound horizon is given by
\ee
r_s(a) = \int_0^a \rd a \frac{\rd \tau}{\rd a} c_s,
\eee
where the speed of sound is $c_s^{-2} = 3 ( 1+R^{-1})$. Here
$R=\frac{4}{3}\frac{\rho_{\gamma}}{\rho_{b}}$ is the photon to baryon
ratio. At redshifts where the speed of sound is appreciable and
$r_s(a)$ receives contributions, $\od(a) \approx \ode$ holds very well
and we get again from the Friedmann Equation that
\ee
r_s = \frac{\sqrt{1-\ode}}{H_0} \int \frac{\rd a}{\sqrt{ 3\left  (\om^0 a + \ore\right)\left(1 + R^{-1} \right)} },
\eee
which can be integrated (similarly to \cite{Hu:1994uz}) to  
\begin{multline} \label{eqn::rs}
r_s = \frac{4 \sqrt{1-\ode}}{3 H_0} \sqrt{\frac{\og^0}{\ob^0 \om^0}}\\ \times 
\ln \frac{ \sqrt{ 1 + R^{-1}_{ls}} + \sqrt{R^{-1}_{ls} + R^{-1}_{equ.}} } { 1 + \sqrt{R^{-1}_{equ.}}}
\end{multline}
Here, $R_{ls}$ and $R_{equ.}$ is the photon to baryon ratio as defined above at last scattering and matter-radiation
equality respectively.

Finally, with $r_s$, $\tau_0$ and $\tau_{ls}$ given, the acoustic scale 
\ee
l_A = \pi \frac{\tau_0 - \tau_{ls}}{r_s},
\eee
can easily be computed.

\section{Connection to Existing $w(a)$ Parameterizations}
It may be worthwhile to relate our parameterization to the 
$w(a)$ parameterization of
\cite{Corasaniti:2002vg} 
\begin{multline}
w^{Coras.}(a) = w_0 + (w_m - w_0) \times \frac{1 + \exp(a_c / \Delta) }{1 + \exp( [a_c-a]/\Delta )} \\
\times \frac{1 - \exp( [1-a]/\Delta)}{1 - \exp(1/\Delta)}. \label{eqn::pier}
\end{multline}
This versatile parameterization is characterized by the equation of state of dark energy today $w_0$,
the dark energy equation of state during earlier epochs $w_m$, a cross-over scale factor $a_c$
and a parameter $\Delta$ controlling the rapidity of this cross-over. 
In our case, $w_0$ has the same meaning, and $w_m=0$ always for our parameterization.
We might roughly identify $a_c$ in \eqref{eqn::pier} with \eqref{eqn::crossover}.
Using equations \eqref{eqn::connectw} and \eqref{eqn::param} and working
to leading order in $\ode/\od^0$, we also get an estimate of the width of the transition
\ee\label{eqn::breite}
\Delta_c \approx \left( 1 - 3^{\frac{1}{3w_0}} \right) a_c.
\eee
We caution the reader that the accuracy of these expressions varies 
and hence the relation of  our parameterization to that given in Equation \eqref{eqn::pier}
should be seen as a semi-quantitative statement.

\begin{figure}
\includegraphics[scale=0.30] {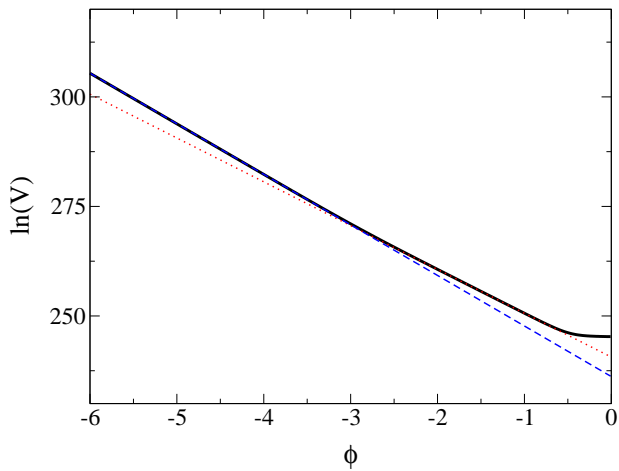}
\caption{The logarithm of the potential, $\ln(V)$ as a function of $\varphi$ in units of $\mplanck$ (solid line).
The exponential potentials during radiation and matter domination are indicated by the dashed (blue)
and dotted (red) line. The exponent in these cases is given by Equation \eqref{eqn::expo}. For this plot,
we used $\ode = 0.03$ and  matter-radiation equality is at $\varphi_{equ.} = -2.66$. In the recent universe, 
the potential flattens, leading to $w \to -0.99$ which we picked for this plot.
\label{fig::vofphi}
}
\end{figure}
\section{The Case of Scalar Dark Energy}
If one uses our parameterization to describe the evolution of 
an (effective \cite{Doran:2002bc}) 
scalar dark energy field, then the constant $\od \approx \ode$ at high
redshifts implies an exponential potential for the scalar field.  Indeed, the 
value of $\ode$ is that of an attractor solution for the potential \cite{Wetterich:fm}  
\ee\label{eqn::expo}
V(\varphi) = \mplanck^4 \exp\left(- \sqrt{ 3 [1+w_{backg}] / {\ode}  } \,\,\, \varphi \right),
\eee
where $\mplanck$ is the reduced Planck mass and 
$w_{backg}$ is the equation of state of the components 
other than dark energy, i.e. $w_{backg}=1/3$ during radiation
domination and $w_{backg}=0$ during matter domination.
As $\ode=const.$ in our parameterization, we conclude that 
in terms of a scalar field potential, we have four ``phases''. 
The potential  (1) is an exponential potential during radiation
domination, (2) moves over to a slightly less steep  exponential potential
during recombination,
(3) stays with that exponential potential during matter domination 
and at late times (4) the potential flattens or -- equivalently -- the kinetic term 
\cite{Armendariz-Picon:2000dh,Hebecker:2000zb} changes. In Figure \ref{fig::vofphi},
we exemplify this behavior by plotting $\varphi$ in units of  $\mplanck$ vs.  $\ln(V)$. For this
plot, we have normalized  $\varphi$ such that today $\varphi=0$.

\begin{figure}
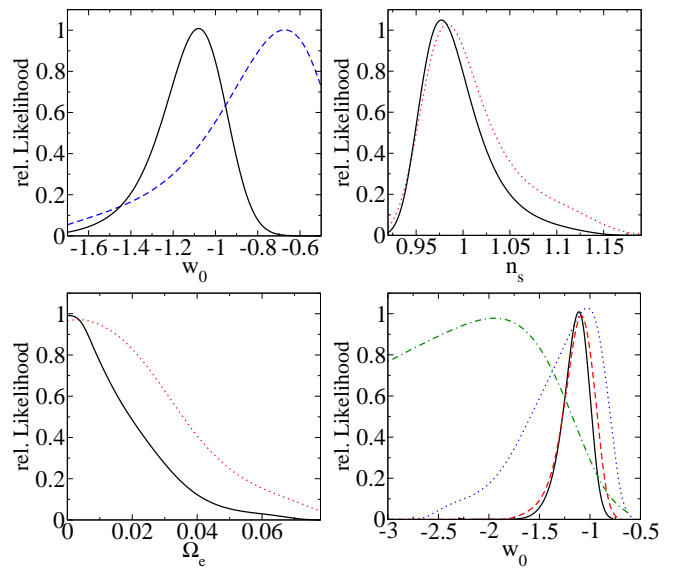

\begin{tabular}{ll}
\includegraphics[height=0.155\textheight,width=0.23\textwidth ]{overlay_phantom_w0}
&\includegraphics[height=0.155\textheight,width=0.23\textwidth ]{overlay_boom_noboom_ns}\\
\includegraphics[height=0.155\textheight,width=0.23\textwidth]{overlay_boom_wmap_oe}
&\includegraphics[height=0.155\textheight,width=0.242\textwidth]{compare_astier_w0}
\end{tabular}
\caption{Upper left panel: Likelihood distribution for the equation of state $w_0$ today for a phantom 
model with $\crest = 1$. The solid line is for WMAP + VSA + CBI +
BOOMERANG + SDSS + SNe Ia, the dashed (blue) line is without Sne Ia
data. At $1\sigma$ confidence, the full data gives % $w_0 =-1.0835 + 0.1441 - 0.1700$.  
$w_0 = -1.08 + 0.14 - 0.17$.
Upper right panel: Likelihood
distribution for the scalar spectral index $n_s$ for a scalar field
model. The solid line is for WMAP + VSA + CBI + BOOMERANG + SDSS + SNe
Ia, the dotted (red) line is without Boomerang. At $1\sigma$
confidence, 
%$n_s = 0.9771 + 0.0373 - 0.0263$
$n_s = 0.98 + 0.04 - 0.03$. 
Lower left panel:
Likelihood distribution for $\ode$ for a scalar field model. The solid
line is for WMAP + VSA + CBI + BOOMERANG + SDSS+SNe Ia, the dotted
(red) line is without Boomerang. 
At $2\sigma$ confidence, $\ode < 0.04$ and $\ode < 0.06$ respectively. 
 Lower right panel: Comparison of the likelihood
distribution for the equation of state $w_0$ today for the gold set
from Riess et. al. only (green, dot-dashed line) and the SNLS data
only (blue, dotted line), and their combination with the CMB+LSS data:
the dashed line (red) corresponds to CMB+LSS+Riess et. al., the solid
(black) line to CMB+LSS+SNLS.
\label{fig::overlay_phantom_w0}
}
\end{figure}

As said, our  parameterization extends to phantom crossing
models.  For $w_0 < -1$,
there is a crossing from the phantom to the ``canonical'' dark energy regime.
In this case, however, it seems impossible to describe the behavior using 
a single scalar field \cite{Vikman:2004dc,Hu:2004kh,Huey:2004jz,Caldwell:2005ai,Zhao:2005vj} .  
For simulations of phantom crossing models, we  chose to fix the rest frame speed of 
sound of dark energy to the speed of light $\crest = 1$, which is compatible with
the ``microscopic'' behavior of scalar dark energy models. Yet,
from the point of view of fundamental physics, it seems ill-suited, because 
the entropy generation rate $\Gamma$  diverges. 
We nevertheless  compute CMB and LSS constraints for phantom crossing models using $c_s^2=1$,
because the detailed choice of $\crest$ seems to be of no significance for
phenomenological studies using current data \cite{Caldwell:2005ai}.

\begin{table*}[!ht] 
\begin{ruledtabular}
\begin{tabular}{ccccccccccc}
& $\Omega_m h^2$ & $\Omega_b h^2$& $h$&$n_s$  & $\tau$ & $\ln(10^{10} A_s) - 2\tau$ &$w_0 $ &$\ode $ \\ \hline \hline
BASE+SNE+B03 &  $0.143^{+0.007}_{-0.007}$ &  $0.023^{+0.001}_{-0.001}$ &  $0.69^{+0.02}_{-0.03}$ &  $0.98^{+0.04}_{-0.03}$  &  $0.13^{ + 0.08}_{ - 0.06}$ &  $2.90^{ + 0.03}_{ - 0.03}$  &$\leq -0.82 $ &$ \leq 0.04$ \\
BASE+SNE       & $0.143^{+0.008}_{-0.008}$ & $0.024^{+0.002}_{-0.001}$ & $0.68^{+0.03}_{-0.03}$ & $0.98^{+0.05}_{-0.03}$ & $0.15^{ + 0.10}_{ - 0.07}$ & $2.89^{ + 0.04}_{ - 0.03}$  &$ \leq -0.82$ &$ \leq 0.06$\\
BASE+B03       & $0.141^{+0.01}_{-0.01}$ & $0.023^{+0.002}_{-0.001}$ & $0.62^{+0.04}_{-0.05}$ & $0.98^{+0.05}_{-0.03}$& $0.13^{ + 0.09}_{ - 0.07}$ & $2.89^{ + 0.04}_{ - 0.03}$ &$ -0.65^{+0.18}_{-0.26}$ &$ \leq 0.04 $ \\
BASE           & $0.142^{+0.01}_{-0.01}$ & $0.024^{+0.002}_{-0.001}$&$0.61^{+0.04}_{-0.05}$&$0.98^{+0.06}_{-0.03}$ & $0.14^{+0.10}_{-0.08}$ & $2.88^{+0.04}_{-0.04}$  &$-0.70^{+0.24}_{-0.28}$ &$ \leq 0.06$ \\ 
 \end{tabular}
\end{ruledtabular}
\caption{Parameter constraints for scalar dark energy models. BASE is WMAP+VSA+CBI+SDSS. All errors are $1\sigma$, upper bounds
are $2\sigma$.}\label{tab::scalar}
\end{table*}

\section{Simulation results}
We added the parameterization \eqref{eqn::param}  to {\sc Cmbeasy} \cite{Doran:2003sy} and computed 
constraints on cosmological parameters using a Markov Chain Monte Carlo 
approach \cite{Lewis:2002ah} with the {\sc AnalyzeThis!} package \cite{Doran:2003ua}. 
The parameters we chose are the matter energy fraction $\omh$,
the baryon energy fraction $\obh$, the hubble parameter $h$, optical depth $\tau$,
scalar spectral index $n_s$, initial scalar amplitude $A_s$ using the observationally relevant 
combination $\ln(10^{10} A_s) - 2\tau$ as well as  the dark energy parameters $w_0$ and $\ode$.
We chose flat priors on all parameters and ran two kinds of models separately: scalar field
dark energy models with $w_0 > -1$  and phantom crossing models with $w_0 \in [-5,0]$ 
and speed of sound $c_s^2 = 1$.
We compare the predictions of the models with several combinations of data sets. 
The first set, CMB+LSS, consists of CMB data from WMAP \cite{Spergel:2003cb},
VSA \cite{Dickinson:2004yr} and CBI \cite{Readhead:2004gy}, as well as LSS data from
SDSS \cite{Tegmark:2003ud}. For SNe Ia, we use the compilation by Riess et. al., \cite{Riess:2004nr}.
Motivated by recent comparisons of different SN Ia data sets \cite{Nesseris:2005ur,Jassal:2006gf},
we also take the first-year data from the SNLS \cite{Astier:2005qq} into
account when considering phantom crossing models.

\begin{table*}[!ht] 
\begin{ruledtabular}
\begin{tabular}{ccccccccccc}
& $\Omega_m h^2$ & $\Omega_b h^2$& $h$&$n_s$  & $\tau$ & $\ln(10^{10} A_s) - 2\tau$ &$w_0 $ &$\ode $ \\ \hline \hline
BASE+B03+SNLS &  $0.15^{+0.01}_{-0.01}$ &  $0.023^{+ 0.001}_{- 0.001}$ &  $0.70^{+0.02}_{- 0.02}$ &  $0.96^{+ 0.03}_{- 0.02}$ &  $0.10^{ +0.07}_{-0.05 }$ &  $2.94^{+ 0.04}_{- 0.04}$ & $-1.11^{+ 0.12 }_{- 0.14} $ &$\leq 0.04 $ \\
BASE+SNLS       & $0.15^{+ 0.01}_{- 0.01}$ & $0.023^{+ 0.001}_{- 0.001}$ & $0.70^{+ 0.03}_{- 0.02}$ & $0.96^{+ 0.03}_{- 0.03}$  &  $0.11^{+ 0.08}_{- 0.05}$ & $2.95^{+ 0.04}_{- 0.04 }$ & $-1.16^{+ 0.13}_{- 0.17}$  &$\leq 0.05 $\\
BASE+SNE        & $0.16^{+ 0.01}_{- 0.01}$ & $0.023^{+ 0.001}_{- 0.001}$ & $0.68^{+ 0.02}_{- 0.02}$ & $0.96^{+ 0.03}_{- 0.02}$ & $0.10^{+ 0.07}_{- 0.05}$ & $2.97^{+ 0.04}_{- 0.05}$& $-1.16^{+ 0.19}_{- 0.23}$ &$\leq 0.06 $ \\
 BASE+B03       & $0.15^{+ 0.01}_{- 0.01}$ & $0.023^{+ 0.001}_{- 0.001}$ & $0.62^{+ 0.07}_{- 0.06}$&$0.96^{+ 0.04}_{- 0.02}$ & $0.11^{+ 0.09}_{- 0.06}$ & $2.92^{ + 0.04}_{ - 0.04}$ &$-0.68^{+ 0.24}_{- 0.32}$ &$\leq 0.04 $ \\
BASE+SNE+B03 & $0.15^{ + 0.01}_{ - 0.01}$&$0.023^{ + 0.001}_{ - 0.001}$&$0.68^{ + 0.02}_{ - 0.02}$&$0.96^{ + 0.03}_{ - 0.02}$ & $0.10^{ + 0.06}_{ - 0.05}$&$2.95^{ + 0.04}_{ - 0.04}$&$-1.08^{ + 0.13}_{ - 0.16}$& $\leq 0.04$ \\
 \end{tabular}
\end{ruledtabular}
\caption{Parameter constraints for phantom crossing models. BASE is WMAP+VSA+CBI+SDSS. All errors are $1\sigma$, upper bounds
are $2\sigma$. }
\label{tab::phantom}
\end{table*}

Turning first to the case of scalar field dark energy, our main results are shown in
Figure \ref{fig::combined_scalar_w0_Oe} and Table \ref{tab::scalar}. We see that the \mbox{SNe Ia} data and
the CMB+LSS data give orthogonal information. While supernovae are insensitive to the amount of
early dark energy $\ode$ but do constrain $w_0$, the opposite is true for the CMB+LSS set.
Figure \ref{fig::combined_scalar_w0_Oe} also shows the likelihood contours obtained when
adding the data from the 2003 flight of Boomerang \cite{Montroy:2005yx, Piacentini:2005yq, Jones:2005yb}.

The stronger constraints on the amount of early dark energy from Boomerang are due
to the particular suppression of power caused by $\ode$: the presence of dark energy suppresses
the growth of linear fluctuations that are inside the horizon. The smaller the scale, the longer
it suffered the suppression. This leads to a scale dependent red tilt of the spectra for all modes
that enter after the scale of equality $k< k_{equ.}$. All modes $k>k_{equ.}$ were inside the
horizon before equality and are suppressed by the same factor \cite{Caldwell:2003vp}.
An increase of $\ode$ is hence
partially degenerate with a decrease of $n_s$. Now Boomerang alone prefers a rather red
spectral index $n_s = 0.86$ \cite{MacTavish:2005yk}, whereas WMAP tends more towards $n_s \sim 1$
(see also the upper right panel of Figure \ref{fig::overlay_phantom_w0}).
As far as early dark energy models with appreciable $\ode$ are concerned, WMAP forces them to
have $n_s$ close to $1$, at least slightly larger than a comparable model with vanishing $\ode$. The relative lack
of power of Boomerang at high multipoles does then disfavor those models with large $\ode$ and rather blue $n_s$.

Considering phantom crossing models, we use the CMB+LSS data, and the
SNe Ia data from either Riess et. al. or the SNLS. In Table \ref{tab::phantom}, we summarize
the constraints. The comparison of
the results for the equation of state parameter today, $w_0$, are
shown in the upper left and lower right panels of Figure
\ref{fig::overlay_phantom_w0}.  We find that the SNLS data alone, in
contrast to the set from Riess et. al., do not favor a value of $w_0 <
1$, which agrees with the findings in \cite{Nesseris:2005ur}. This
situation changes when combining the supernovae data with CMB+LSS,
since the CMB essentially fixes $\omh$. Without a free $\Omega_m^0$,
both supernovae sets prefer a $w_0$ of slightly less than~$-1$
\cite{Alam:2005pb,Xia:2005ge,Ichikawa:2005nb,Cai:2005qm}.

\section{Conclusions}
We introduced a new parameterization of dark energy. It is
particularly well suited to describe models in which the dark energy
density is non-negligible in the early Universe. Working with $\od(a)$
instead of $w(a)$ facilitates the computation of quantities such as
horizons, the luminosity distance and the acoustic scale. Using Markov
Chain Monte Carlo simulations, we constrained the parameters of our
model using the latest CMB, Sne Ia and LSS data.  While the CMB is more
sensitive to $\ode$, the opposite is true for Sne Ia, which are more
sensitive to $w_0$. Adding the recent high multipole Boomerang data
tightened the upper bound on $\ode$ considerably to $\ode < 0.04$. However,
our analysis only included linear fluctuations.  As early dark energy
models lead to more non-linear structure at higher redshifts compared
to a cosmological constant \cite{Bartelmann:2005fc}, it might well be
that the upper bound turns into a detection in the future.

{\bf Acknowledgments} 
We would like to thank Ben Gold for his helpful comments on the preprint.
%%%%%%%%%%%%%%%%%%%%

%%%%%%%%%%%%%%%%%%%%

\section{Appendix}
The upper bound for $\ode$ with our parameterization in the scalar dark energy case is
considerably higher than the one found in \cite{Doran:2005sn}.
This is a  consequence of the longer period of time for which our
parameterization mimics a cosmological constant, so that the influence
of $\ode$ on the late-time evolution of the Universe induced by the
choice of parameterization is minimized. It is then natural to ask
if the bounds on $\ode$ in our model might relax if
$\od(a)$ behaved like a cosmological constant for higher redshifts. 
To answer this question, we  generalize our
parameterization  Equation \eqref{eqn::param} by introducing
an additional parameter $\gamma$ that controls the importance of the terms
involving $\ode$ at late times,
\ee \label{eq::extparam}
\od(a) =  \frac{\od^0 - \ode \left(1- a^{-3w_0}\right)^\gamma}{\od^0 + \om^0 a^{3w_0}} + \ode \left (1- a^{-3w_0}\right)^\gamma.
\eee
When setting $\gamma = 1$, this reduces to our parameterization
\eqref{eqn::param}, while increasing $\gamma$ shifts the departure
from a cosmological constant-like behavior of $\od(a)$ to higher
redshifts. If we require $\od(a)$ to stay monotonic, then $\gamma$ is
bounded by
\ee
\gamma \lesssim \frac{\od-\ode}{\ode(1-\od)}.
\eee
Repeating the Monte-Carlo analysis for the scalar dark energy case
with this extended parameterization \eqref{eq::extparam} at different values
of $\gamma$ does not alter the constraints on the other parameters.
It  therefore suffices to use the simpler parameterization of
Equation \eqref{eqn::param}.


\begin{thebibliography}{}

%\cite{Astier:2005qq}
\bibitem{Astier:2005qq}
  P.~Astier {\it et al.},
  %``The Supernova Legacy Survey: Measurement of Omega_M, Omega_Lambda and w
  %from the First Year Data Set,''
  arXiv:astro-ph/0510447.
  %%CITATION = ASTRO-PH 0510447;%%

%\cite{Riess:2004nr}
\bibitem{Riess:2004nr}
A.~G.~Riess {\it et al.}  [Supernova Search Team Collaboration],
%``Type Ia Supernova Discoveries at z>1 From the Hubble Space Telescope:
%Evidence for Past Deceleration and Constraints on Dark Energy Evolution,''
Astrophys.\ J.\  {\bf 607}, 665 (2004)
[arXiv:astro-ph/0402512].
%%CITATION = ASTRO-PH 0402512;%%

%\cite{Spergel:2003cb}
\bibitem{Spergel:2003cb}
D.~N.~Spergel {\it et al.}  [WMAP Collaboration],
%``First Year Wilkinson Microwave Anisotropy Probe (WMAP) Observations:
%Determination of Cosmological Parameters,''
Astrophys.\ J.\ Suppl.\  {\bf 148}, 175 (2003)
[arXiv:astro-ph/0302209].
%%CITATION = ASTRO-PH 0302209;%%

%\cite{Readhead:2004gy}
\bibitem{Readhead:2004gy}
A.~C.~S.~Readhead {\it et al.},
%``Extended Mosaic Observations with the Cosmic Background Imager,''
Astrophys.\ J.\  {\bf 609} (2004) 498
[arXiv:astro-ph/0402359].
%%CITATION = ASTRO-PH 0402359;%%

%\cite{Dickinson:2004yr}
\bibitem{Dickinson:2004yr}
  C.~Dickinson {\it et al.},
  %``High sensitivity measurements of the CMB power spectrum with the extended
  %Very Small Array,''
  Mon.\ Not.\ Roy.\ Astron.\ Soc.\  {\bf 353} (2004) 732
  [arXiv:astro-ph/0402498].
  %%CITATION = ASTRO-PH 0402498;%%

%\cite{Tegmark:2003ud}
\bibitem{Tegmark:2003ud}
M.~Tegmark {\it et al.}  [SDSS Collaboration],
%``Cosmological parameters from SDSS and WMAP,''
Phys.\ Rev.\ D {\bf 69} (2004) 103501
[arXiv:astro-ph/0310723].
%%CITATION = ASTRO-PH 0310723;%%

%\cite{Wetterich:fm}
\bibitem{Wetterich:fm}
C.~Wetterich,
%``Cosmology And The Fate Of Dilatation Symmetry,''
Nucl.\ Phys.\ B {\bf{302}}, 668  (1988)
%%CITATION = NUPHA,B302,668;%%

%\cite{Ratra:1987rm}
\bibitem{Ratra:1987rm}
B.~Ratra and P.~J.~Peebles,
%``Cosmological Consequences Of A Rolling Homogeneous Scalar Field,''
Phys.\ Rev.\ D {\bf{37}}, 3406  (1988)
%%CITATION = PHRVA,D37,3406;%%

%\cite{Caldwell:1997ii}
\bibitem{Caldwell:1997ii}
R.~R.~Caldwell,~R.~Dave and P.~J.~Steinhardt,
%``Cosmological Imprint of an Energy Component with General Equation-of-State,''
Phys.\ Rev.\ Lett.\  {\bf{80}}, 1582 (1998)
%[preprint (astro-ph/9708069].
%%CITATION = ASTRO-PH 9708069;%%

%\cite{Cooray:1999da}
\bibitem{Cooray:1999da}
  A.~R.~Cooray and D.~Huterer,
  %``Gravitational Lensing as a Probe of Quintessence,''
  Astrophys.\ J.\  {\bf 513}, L95 (1999)
  [arXiv:astro-ph/9901097].
  %%CITATION = ASTRO-PH 9901097;%%

%\cite{Huterer:2000mj}
\bibitem{Huterer:2000mj}
  D.~Huterer and M.~S.~Turner,
  %``Probing the dark energy: Methods and strategies,''
  Phys.\ Rev.\ D {\bf 64}, 123527 (2001)
  [arXiv:astro-ph/0012510].
  %%CITATION = ASTRO-PH 0012510;%%

%\cite{Corasaniti:2002vg}
\bibitem{Corasaniti:2002vg}
P.~S.~Corasaniti and E.~J.~Copeland,
%``A model independent approach to the dark energy equation of state,''
Phys.\ Rev.\ D {\bf 67}, 063521 (2003)
[arXiv:astro-ph/0205544].
%%CITATION = ASTRO-PH 0205544;%%

\bibitem{Linder:2002et}
E.~V.~Linder,
%``Exploring the expansion history of the universe,''
Phys.\ Rev.\ Lett.\  {\bf 90}, 091301 (2003).
%[arXiv:astro-ph/0208512].



%\cite{Chevallier:2000qy}
\bibitem{Chevallier:2000qy}
  M.~Chevallier and D.~Polarski,
  %``Accelerating universes with scaling dark matter,''
  Int.\ J.\ Mod.\ Phys.\ D {\bf 10}, 213 (2001)
  [arXiv:gr-qc/0009008].
  %%CITATION = GR-QC 0009008;%%

%\cite{Upadhye:2004hh}
\bibitem{Upadhye:2004hh}
  A.~Upadhye, M.~Ishak and P.~J.~Steinhardt,
  %``Dynamical dark energy: Current constraints and forecasts,''
  arXiv:astro-ph/0411803.
  %%CITATION = ASTRO-PH 0411803;%%

%\cite{Wang:2004py}
\bibitem{Wang:2004py}
  Y.~Wang and M.~Tegmark,
  %``New dark energy constraints from supernovae, microwave background and
  %galaxy clustering,''
  Phys.\ Rev.\ Lett.\  {\bf 92}, 241302 (2004)
  [arXiv:astro-ph/0403292].
  %%CITATION = ASTRO-PH 0403292;%%

%\cite{Doran:2000jt}
\bibitem{Doran:2000jt}
M.~Doran, M.~J.~Lilley, J.~Schwindt and C.~Wetterich,
%``Quintessence and the separation of CMB peaks,''
Astrophys.\ J.\  {\bf 559}, 501 (2001)
[arXiv:astro-ph/0012139].
%%CITATION = ASTRO-PH 0012139;%%

%\cite{Doran:2001rw}
\bibitem{Doran:2001rw}
M.~Doran, J.~M.~Schwindt and C.~Wetterich,
%``Structure formation and the time dependence of quintessence,''
Phys.\ Rev.\ D {\bf 64}, 123520 (2001)
[arXiv:astro-ph/0107525].
%%CITATION = ASTRO-PH 0107525;%%

%\cite{Bartelmann:2005fc}
\bibitem{Bartelmann:2005fc}
  M.~Bartelmann, M.~Doran and C.~Wetterich,
  %``Non-linear Structure Formation in Cosmologies with Early Dark Energy,''
  arXiv:astro-ph/0507257.
  %%CITATION = ASTRO-PH 0507257;%%

\bibitem{Wetterich:1994bg}
  C.~Wetterich,
  %``The Cosmon model for an asymptotically vanishing time dependent
  %cosmological 'constant',''
  Astron.\ Astrophys.\  {\bf 301}, 321 (1995)
  [arXiv:hep-th/9408025].
  %%CITATION = HEP-TH 9408025;%%



%\cite{Bean:2001wt}
\bibitem{Bean:2001wt}
  R.~Bean, S.~H.~Hansen and A.~Melchiorri,
  %``Early universe constraints on dark energy,''
  Phys.\ Rev.\ D {\bf 64}, 103508 (2001)
  [arXiv:astro-ph/0104162].
  %%CITATION = ASTRO-PH 0104162;%%

%\cite{Muller:2004gu}
\bibitem{Muller:2004gu}
  C.~M.~Muller, G.~Schafer and C.~Wetterich,
  %``Nucleosynthesis and the variation of fundamental couplings,''
  Phys.\ Rev.\ D {\bf 70}, 083504 (2004)
  [arXiv:astro-ph/0405373].
  %%CITATION = ASTRO-PH 0405373;%%

%\cite{Wetterich:2004pv}
\bibitem{Wetterich:2004pv}
C.~Wetterich,
%``Phenomenological parameterization of quintessence,''
Phys.\ Lett.\ B {\bf 594}, 17 (2004)
[arXiv:astro-ph/0403289].
%%CITATION = ASTRO-PH 0403289;%%

%\cite{Feng:2004ad}
\bibitem{Feng:2004ad}
  B.~Feng, X.~L.~Wang and X.~M.~Zhang,
  %``Dark Energy Constraints from the Cosmic Age and Supernova,''
  Phys.\ Lett.\ B {\bf 607}, 35 (2005)
  [arXiv:astro-ph/0404224].
  %%CITATION = ASTRO-PH 0404224;%%

%\cite{Vikman:2004dc}
\bibitem{Vikman:2004dc}
  A.~Vikman,
  %``Can dark energy evolve to the phantom?,''
  Phys.\ Rev.\ D {\bf 71}, 023515 (2005)
  [arXiv:astro-ph/0407107].
  %%CITATION = ASTRO-PH 0407107;%%

%\cite{Hu:2004kh}
\bibitem{Hu:2004kh}
  W.~Hu,
  %``Crossing the phantom divide: Dark energy internal degrees of freedom,''
  Phys.\ Rev.\ D {\bf 71} (2005) 047301
  [arXiv:astro-ph/0410680].
  %%CITATION = ASTRO-PH 0410680;%%

%\cite{Huey:2004jz}
\bibitem{Huey:2004jz}
  G.~Huey,
  %``A Comprehensive Approach to Resolving the Nature of the Dark Energy,''
  arXiv:astro-ph/0411102.
  %%CITATION = ASTRO-PH 0411102;%%

%\cite{Caldwell:2005ai}
\bibitem{Caldwell:2005ai}
  R.~R.~Caldwell and M.~Doran,
  %``Dark-energy evolution across the cosmological-constant boundary,''
  Phys.\ Rev.\ D {\bf 72} (2005) 043527
  [arXiv:astro-ph/0501104].
  %%CITATION = ASTRO-PH 0501104;%%

%\cite{Zhao:2005vj}
\bibitem{Zhao:2005vj}
  G.~B.~Zhao, J.~Q.~Xia, M.~Li, B.~Feng and X.~Zhang,
  %``Perturbations of the quintom models of dark energy and the effects on
  %observations,''
  Phys.\ Rev.\ D {\bf 72}, 123515 (2005)
  [arXiv:astro-ph/0507482].
  %%CITATION = ASTRO-PH 0507482;%%


%\cite{Gruppuso:2005xy}
\bibitem{Gruppuso:2005xy}
  A.~Gruppuso and F.~Finelli,
  %``Analytic results for a flat universe dominated by dust and dark energy,''
  arXiv:astro-ph/0512641.
  %%CITATION = ASTRO-PH 0512641;%%

%\cite{Hu:1994uz}
\bibitem{Hu:1994uz}
  W.~Hu and N.~Sugiyama,
  %``Anisotropies in the Cosmic Microwave Background: An Analytic Approach,''
  Astrophys.\ J.\  {\bf 444} (1995) 489
  [arXiv:astro-ph/9407093].
  %%CITATION = ASTRO-PH 9407093;%%

%\cite{Doran:2002bc}
\bibitem{Doran:2002bc}
  M.~Doran and J.~Jaeckel,
  %``Loop corrections to scalar quintessence potentials,''
  Phys.\ Rev.\ D {\bf 66}, 043519 (2002)
  [arXiv:astro-ph/0203018].
  %%CITATION = ASTRO-PH 0203018;%%
  %%Cited 27 times in SPIRES-HEP

%\cite{Armendariz-Picon:2000dh}
\bibitem{Armendariz-Picon:2000dh}
  C.~Armendariz-Picon, V.~Mukhanov and P.~J.~Steinhardt,
  %``A dynamical solution to the problem of a small cosmological constant  and
  %late-time cosmic acceleration,''
  Phys.\ Rev.\ Lett.\  {\bf 85}, 4438 (2000)
  [arXiv:astro-ph/0004134].
  %%CITATION = ASTRO-PH 0004134;%%

%\cite{Hebecker:2000zb}
\bibitem{Hebecker:2000zb}
  A.~Hebecker and C.~Wetterich,
  %``Natural quintessence?,''
  Phys.\ Lett.\ B {\bf 497} (2001) 281
  [arXiv:hep-ph/0008205].
  %%CITATION = HEP-PH 0008205;%%

%\cite{Doran:2003sy}
\bibitem{Doran:2003sy}
  M.~Doran,
  %``CMBEASY:: an Object Oriented Code for the Cosmic Microwave Background,''
  JCAP {\bf 0510}, 011 (2005)
  [arXiv:astro-ph/0302138].
  %%CITATION = ASTRO-PH 0302138;%%

%\cite{Lewis:2002ah}
\bibitem{Lewis:2002ah}
  A.~Lewis and S.~Bridle,
  %``Cosmological parameters from CMB and other data: a Monte-Carlo approach,''
  Phys.\ Rev.\ D {\bf 66} (2002) 103511
  [arXiv:astro-ph/0205436].
  %%CITATION = ASTRO-PH 0205436;%%

%\cite{Doran:2003ua}
\bibitem{Doran:2003ua}
  M.~Doran and C.~M.~Mueller,
  %``Analyze This! A Cosmological Constraint Package for cmbeasy,''
  JCAP {\bf 0409}, 003 (2004)
  [arXiv:astro-ph/0311311].
  %%CITATION = ASTRO-PH 0311311;%%

%\cite{Nesseris:2005ur}
\bibitem{Nesseris:2005ur}
S.~Nesseris and L.~Perivolaropoulos,
%``Comparison of the Legacy and Gold SnIa Dataset Constraints on Dark Energy
%Models,''
arXiv:astro-ph/0511040.
%%CITATION = ASTRO-PH 0511040;%%

%\cite{Jassal:2006gf}
\bibitem{Jassal:2006gf}
H.~K.~Jassal, J.~S.~Bagla and T.~Padmanabhan,
%``The vanishing phantom menace,''
arXiv:astro-ph/0601389.
%%CITATION = ASTRO-PH 0601389;%%

%\cite{Montroy:2005yx}
\bibitem{Montroy:2005yx}
T.~E.~Montroy {\it et al.},
%``A Measurement of the CMB  Spectrum from the 2003 Flight of BOOMERANG,''
arXiv:astro-ph/0507514.
%%CITATION = ASTRO-PH 0507514;%%

%\cite{Piacentini:2005yq}
\bibitem{Piacentini:2005yq}
F.~Piacentini {\it et al.},
%``A measurement of the polarization-temperature angular cross power spectrum of
%the Cosmic Microwave Background from the 2003 flight of BOOMERANG,''
arXiv:astro-ph/0507507.
%%CITATION = ASTRO-PH 0507507;%%

%\cite{Jones:2005yb}
\bibitem{Jones:2005yb}
W.~C.~Jones {\it et al.},
%``A Measurement of the Angular Power Spectrum of the CMB Temperature Anisotropy
%from the 2003 Flight of Boomerang,''
arXiv:astro-ph/0507494.
%%CITATION = ASTRO-PH 0507494;%%

%\cite{Caldwell:2003vp}
\bibitem{Caldwell:2003vp}
R.~R.~Caldwell, M.~Doran, C.~M.~Mueller, G.~Schaefer and C.~Wetterich,
%``Early quintessence in light of WMAP,''
Astrophys.\ J.\  {\bf 591} (2003) L75
[arXiv:astro-ph/0302505].
%%CITATION = ASTRO-PH 0302505;%%

%\cite{MacTavish:2005yk}
\bibitem{MacTavish:2005yk}
  C.~J.~MacTavish {\it et al.},
  %``Cosmological parameters from the 2003 flight of BOOMERANG,''
  arXiv:astro-ph/0507503.
  %%CITATION = ASTRO-PH 0507503;%%

%\cite{Alam:2005pb}
\bibitem{Alam:2005pb}
U.~Alam and V.~Sahni,
%``Confronting braneworld cosmology with supernova data and baryon
%oscillations,''
arXiv:astro-ph/0511473.
%%CITATION = ASTRO-PH 0511473;%%

%\cite{Xia:2005ge}
\bibitem{Xia:2005ge}
J.~Q.~Xia, G.~B.~Zhao, B.~Feng, H.~Li and X.~Zhang,
%``Observing dark energy dynamics with supernova, microwave background and
%galaxy clustering,''
arXiv:astro-ph/0511625.
%%CITATION = ASTRO-PH 0511625;%%

%\cite{Ichikawa:2005nb}
\bibitem{Ichikawa:2005nb}
K.~Ichikawa and T.~Takahashi,
%``Dark energy evolution and the curvature of the universe from recent
%observations,''
arXiv:astro-ph/0511821.
%%CITATION = ASTRO-PH 0511821;%%

%\cite{Cai:2005qm}
\bibitem{Cai:2005qm}
R.~G.~Cai, Y.~Gong and B.~Wang,
%``Super-acceleration on the brane by energy flow from the bulk,''
arXiv:hep-th/0511301.
%%CITATION = HEP-TH 0511301;%%

%\cite{Doran:2005sn}
\bibitem{Doran:2005sn}
M.~Doran, K.~Karwan and C.~Wetterich,
%``Observational constraints on the dark energy density evolution,''
JCAP {\bf 0511} (2005) 007
[arXiv:astro-ph/0508132].
%%CITATION = ASTRO-PH 0508132;%%
  %%Cited 30 times in SPIRES-HEP
\end{thebibliography}
\end{document}